\documentclass[prd,twocolumn,amsmath,amssymb,nofootinbib,preprintnumbers,balancelastpage]{revtex4}
\usepackage{graphicx}
\usepackage{hyperref}
\hypersetup{
    pdfnewwindow=true,      
    colorlinks=true,        
    linkcolor=black,        
    citecolor=blue,         
    filecolor=blue,         
    urlcolor=blue           
}
\begin{document}
\title{\Large {\bf{Baryon Asymmetry, Dark Matter and Local Baryon Number}}}
\author{Pavel Fileviez P\'erez}
\email{fileviez@mpi-hd.mpg.de}
\affiliation{\vspace{0.15cm} \\  Particle and Astro-Particle Physics Division \\
Max-Planck Institute for Nuclear Physics {\rm{(MPIK)}} \\
Saupfercheckweg 1, 69117 Heidelberg, Germany}
\author{Hiren H. Patel}
\email{hiren.patel@mpi-hd.mpg.de}
\affiliation{\vspace{0.15cm} \\  Particle and Astro-Particle Physics Division \\
Max-Planck Institute for Nuclear Physics {\rm{(MPIK)}} \\
Saupfercheckweg 1, 69117 Heidelberg, Germany}

\begin{abstract}
We propose a new mechanism to understand the relation between baryon and dark matter asymmetries in the universe in theories where the baryon number 
is a local symmetry. In these scenarios the $B-L$ asymmetry generated through a mechanism such as leptogenesis is transferred to the dark matter 
and baryonic sectors through sphalerons processes which conserve total baryon number. We show that it is possible to have a consistent 
relation between the dark matter relic density and the baryon asymmetry in the universe even if the baryon number is broken at the low scale through the Higgs mechanism. We also discuss the case where one uses the Stueckelberg mechanism to understand the conservation of baryon number in nature.
\end{abstract}

\maketitle
\section{Introduction}
The existence of the baryon asymmetry and cold dark matter density in the Universe 
has motivated many studies in cosmology and particle physics. Recently, there has been focus on investigations of possible mechanisms that relate the baryon asymmetry and dark matter density, i.e. $\Omega_{DM} \sim 5 \Omega_{B}$. These mechanisms 
are based on the idea of asymmetric dark matter (ADM). See Ref.~\cite{Zurek} for a recent review 
of ADM mechanisms and Ref.~\cite{Michael} for a review on Baryogenesis at the low scale.

Crucial to baryogenesis in any theory is the existence of baryon violating processes to generate the observed baryon asymmetry in the early universe. In theories where the baryon number is a local symmetry and spontaneously broken at low scale one could have a mechanism which explains the 
relation of the baryon asymmetry and dark matter. For simple realistic theories where the baryon number is a local symmetry, see 
Refs.~\cite{paper1,paper2,paper3}, and for studies of baryogenesis in this context, see e.g. Ref.~\cite{Dulaney}. 

\begin{figure}[t] 
   \vspace{10mm}
   \includegraphics[width=2.2in]{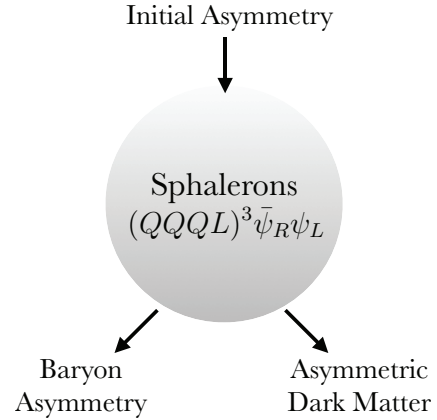} 
   \caption{Schematic representation of the simultaneous generation of the baryon and dark matter asymmetry through the sphalerons.}
   \label{fig:paradigm}
\end{figure}

In this article, we propose a new mechanism to simultaneously generate the baryon and dark matter asymmetries through the electroweak sphalerons.  The basic paradigm is illustrated in Fig.~\ref{fig:paradigm}.  In this model baryon number is a local gauge symmetry which is spontaneously broken at the low scale.  Therefore, the sphalerons should conserve total baryon number.  In this context, the baryonic anomalies in the standard model are cancelled by adding a simple set of vector-like fermionic fields with different baryon numbers.  Furthermore, one can generate masses for all new fields through the Higgs mechanism consistent with collider bounds.  This simple theory provides a fermionic dark matter candidate for which stability is a natural consequence of the spontaneous breaking of baryon number at the low scale. See Ref.~\cite{paper3} for details.

We perform an equilibrium thermodynamic analysis that connects the chemical potentials of the standard model quarks and of the dark matter candidate.  In this way, we find a consistent connection between baryon asymmetry and dark matter relic density.  For completeness, we perform the study in three possible cases.  In the first case, baryon number is conserved in nature, and the leptophobic gauge boson acquires its mass through the Stueckelberg mechanism.  In the second and third cases, baryon number is broken through the Higgs mechanism. However, only in the third case one can have a Higgs boson with the properties consistent with the recent observations at collider experiments.

This article is organized as follows: In section II we define the theory where baryon number is a local gauge 
symmetry, and we discuss the different mechanisms for generating the leptophobic gauge boson mass. In section 
III we perform the equilibrium thermodynamic analysis and find the connection between the baryon asymmetry and dark 
matter density.  Finally, in section IV we discuss the main results.
\section{Local Baryon Number}
In order to understand the conservation or violation of Baryon number in nature we can study a simple model where baryon number is a local symmetry.
The theory is based on the gauge group
 \begin{equation*}
  SU(3)_C \otimes SU(2)_L \otimes U(1)_Y \otimes U(1)_B .
 \end{equation*}
An anomaly-free theory is defined by including additional fermions that account for anomaly cancellation.  A simple set of fields is given by 
\begin{gather}
\begin{aligned}
\Psi_L  &\sim  (1, 2, -\frac{1}{2}, B_1),\\
\eta_R  &\sim  (1, 1, -1, B_1),\\
\chi_R &\sim (1, 1, 0, B_1), \hspace{3mm} \text{and}
\end{aligned}\hspace{3mm}
\begin{aligned}
\Psi_R &\sim (1, 2, -\frac{1}{2}, B_2), \\
\eta_L &\sim  (1, 1, -1, B_2), \\ 
\chi_L &\sim  (1, 1, 0, B_2)\,,
\end{aligned}
\end{gather}
where we use the convention that $Q = T_3 + Y$ to define the normalization of hypercharge. 
Requirement of anomaly cancellation implies the relation
\begin{equation}\label{eq:B}
B_1 - B_2 = -3,
\end{equation}
on the assignment of baryon number to the new fermions.

Notice that the baryon numbers of the vector-like fields must be different. Since the local baryon number is an 
abelian symmetry we can generate mass for the new gauge boson $Z_B$ associated with $U(1)_B$ with or without the Higgs mechanism.  
Therefore, we consider here three different cases:
\begin{itemize}

\item Scenario 1:  The Stueckelberg mechanism generates mass for the leptophobic gauge boson $Z_B$, and the new fermions acquire mass only from 
the standard model (SM) Higgs boson.
See for example Ref.~\cite{Nath} for the application of the Stueckelberg mechanism in this context. 
Unfortunately, because of its sizable couplings to the new fermions, this scenario predicts a reduced branching fraction of the Higgs decay to two photons that is at variance with collider experiments.  Of course, one can add a set of scalar fields with negative couplings to offset the diphoton branching fraction.  But we do not see this possibility as very appealing. See Ref.~\cite{Koji} for the study of the Higgs decays in theories with extra chiral fermions.

\item Scenario 2: Using the Higgs mechanism with a new scalar field $S_B$ with baryon number assignment different from $\pm3$ we can generate mass only for the gauge boson $Z_B$.  As in the first scenario one will run into problems with the predictions of the SM Higgs decays.  It is important to mention that these two scenarios are equivalent in the case when the new Higgs is heavy.

\item Scenario 3: The third scenario corresponds to the case when a new scalar field $S_B$ has baryon number $\pm3$.  This allows for vector-like masses for all new fermionic fields in the theory.  This scenario is consistent with bounds from colliders.
\end{itemize}

It is important to mention that the idea of using the Higgs mechanism to understand the breaking of local baryon number was proposed by A. Pais in Ref.~\cite{Pais}. See also Refs.~\cite{Pati,Frampton} for other theories defined at the high scale where baryon number is conserved. For previous studies where the dark matter candidate has baryon number see Refs.~\cite{BDM1,BDM2,Duerr}.

In this article we will investigate scenario 3 in detail since we can have a consistent model with cosmology and particle physics. We will also present the final result for the first two scenarios.

{\textit{Symmetry Breaking and Mass Generation}}: In order to generate masses in a consistent way we extend the scalar sector with a new Higgs boson $S_B$ to allow for spontaneous breaking of baryon number, 
\begin{eqnarray}
S_{B} & \sim & (1, 1, 0, -3). 
\end{eqnarray}
Notice that the baryon number of the new scalar field is determined once we impose the condition that we need to generate vector-like masses for the new fermions. 
The relevant interactions of the new fields are 
\begin{eqnarray}
-\mathcal{L} &\supset&  Y_1 \bar{\Psi}_L H \eta_R \ + \ Y_2 \bar{\Psi}_R H \eta_L \ + \ Y_3 \bar{\Psi}_L \tilde{H} \chi_R \nonumber \\
& + & Y_4 \bar{\Psi}_R \tilde{H} \chi_L \ + \  \lambda_1 \bar{\Psi}_L \Psi_R S_{B} + \lambda_2 \bar{\eta}_R \eta_L S_{B} \nonumber \\
&+ & \lambda_3 \bar{\chi}_R \chi_L S_{B} + \text{h.c.}
\end{eqnarray}
For the special case $B_1=-B_2$, terms such as $a_1 \chi_L \chi_L S_{B}$ and $a_2 \chi_R \chi_R S_{B}^\dagger$ are allowed.  However, we will focus on the more generic case where $B_1 \neq - B_2$ and where the dark matter annihilation through $Z_B$ is not velocity suppressed~\cite{Duerr}.

Once the new Higgs $S_B$ acquires a vacuum expectation value spontaneously breaking local $U(1)_B$, no operator mediating proton decay is generated, and the scale for baryon number violation can be very low.   For the bounds on the mass of a leptophobic neutral gauge boson see Refs.~\cite{Dobrescu:2013cmh,An:2012ue}. 

The model enjoys three anomaly-free symmetries:

\begin{itemize}

\item $B-L$ in the Standard Model sector is conserved in the usual way. We will refer to this symmetry as $(B-L)_{SM}$.  

\item The accidental $\eta$ global symmetry in the new sector. Under this symmetry the new fields transform in the following way:
\begin{align*}
\Psi_{L,R} \to e^{i \eta}  \Psi_{L,R}, \\ 
 \eta_{L,R} \to e^{i \eta} \eta_{L,R}, \\
 \chi_{L,R} \to e^{i \eta} \chi_{L,R}. 
\end{align*}
This conserved charge will determine the dark matter asymmetry as pointed out in models with vector-like fermions in Ref.~\cite{ADM-vector}. 
Notice that this symmetry guarantees the stability of the lightest field in the new sector.  Therefore, when this new field is electrically neutral, one can have a candidate for cold dark matter in the Universe. 

\item Above the symmetry breaking scale, total baryon number $B_T$ carried by particles in both, the standard model and the new sector is also conserved. Here we will assume that the symmetry breaking scale for baryon number is low, close to the electroweak scale. 

\end{itemize}
We shall specialize to the case when the dark matter is SM singlet-like and is a Dirac fermion $\chi = \chi_L + \chi_R$. 
The interaction relevant for dark matter annihilation into SM quarks via $Z_B$ is given by
\begin{equation}
\mathcal{L} \supset g_B \bar{\chi} \gamma_\mu  Z_B^\mu \left( B_2 P_L + B_1 P_R \right) \chi,
\end{equation}
where $P_L$ and $P_R$ are the left- and right-handed projectors, and $g_B$ is the $U(1)_B$ gauge coupling constant. 
We now proceed to derive the relationship between the baryon asymmetry and dark matter density.
\section{Baryon and Dark Matter Asymmetries}
We recall the thermodynamic relationship between the number density asymmetry $n_+ - n_-$ and the chemical potential
in the limit $\mu \ll T$:
\begin{equation}
\frac{\Delta n}{s}=\frac{n_+ - n_-}{s}=\frac{15 g}{2\pi^2 g_{*} \xi} \frac{\mu}{T},
\end{equation}
where $g$ counts the internal degrees of freedom, $s$ is the entropy density, and $g_{*}$ 
is the total number of relativistic degrees of freedom. Here $\xi=2$ for fermions and $\xi=1$ for bosons.

In order to derive the relationship between the baryon asymmetry and the dark matter relic density, we define the densities associated with the conserved charges in this theory.  Following the notation of Ref.~\cite{Harvey}, the $B-L$ asymmetry in the SM sector is defined in the usual way
\begin{eqnarray}
\Delta (B-L)_{SM} &=& \frac{15}{ 4 \pi^2 g_{*} T}   3 ( \mu_{u_L} +  \mu_{u_R} +  \mu_{d_L}  + \mu_{d_R}  \nonumber \\
&&\qquad-  \mu_{\nu_L} -  \mu_{e_L} -  \mu_{e_R}),
\end{eqnarray}
and the $\eta$ charge density is given by
\begin{equation}
\begin{aligned}
\Delta \eta  &= \frac{15}{ 4 \pi^2 g_{*} T}   \big( 2 \mu_{\Psi_L} +  2 \mu_{\Psi_R} \\
&\qquad +  \mu_{\chi_L}  + \mu_{\chi_R}  +  \mu_{\eta_L} +  \mu_{\eta_R} \big).
\end{aligned}
\end{equation}
Isospin conservation $T_3=0$ implies
\begin{equation}
\mu_{u_L}=\mu_{d_L},\qquad \mu_{e_L}=\mu_{\nu_L}, \qquad \mu_+ = \mu_0\,.
\end{equation}
Standard Model interactions give us the following equilibrium conditions for the chemical potentials~\cite{Harvey}:
\begin{equation}
\begin{aligned}
\mu_{u_R}&=\mu_0 + \mu_{u_L}, \\
\mu_{d_R}&= - \mu_0 + \mu_{u_L},  \\   
\mu_{e_R}&= - \mu_0 + \mu_{e_L}.
\end{aligned}
\end{equation}
The new $\lambda_i$ interactions implies
\begin{equation}
\begin{aligned}
 - \mu_{\Psi_L} + \mu_{\Psi_R} + \mu_{S_B}&=0, \\
 - \mu_{\chi_R} + \mu_{\chi_L} + \mu_{S_B}&=0, \\
 - \mu_{\eta_R} + \mu_{\eta_L} + \mu_{S_B}&=0,
\end{aligned}
\end{equation}
while the $Y_i$ couplings give us
\begin{equation}
\begin{aligned}
 - \mu_{\Psi_L} - \mu_{0} + \mu_{\chi_R}&=0, \\
 - \mu_{\Psi_L} + \mu_{0} + \mu_{\eta_R}&=0, \\
 - \mu_{\Psi_R} - \mu_{0} + \mu_{\chi_L}&=0, \\
 - \mu_{\Psi_R} + \mu_{0} + \mu_{\eta_L}&=0.
\end{aligned}
\end{equation}
Conservation of electric charge $Q_{em}=0$ implies
\begin{multline}
6 (\mu_{u_L} + \mu_{u_R}) - 3 (\mu_{d_L} + \mu_{d_R}) - 3 (\mu_{e_L} + \mu_{e_R}) +   \\
2 \mu_0 -  \mu_{\Psi_L}- \mu_{\Psi_R} -  \mu_{\eta_L}- \mu_{\eta_R}=0. 
\end{multline}
Electroweak sphalerons satisfies the conservation of total baryon number, and the associated `t Hooft operator is
\begin{equation}
(QQQL)^3\bar\psi_R\psi_L\,.
\end{equation}
Therefore, the sphalerons give us an additional equilibrium condition between the standard model particles and new degrees of freedom
\begin{equation}
3 ( 3 \mu_{u_L} + \mu_{e_L}) + \mu_{\Psi_L} - \mu_{\Psi_R}=0. 
\end{equation}
Notice that the sphalerons play the crucial role of transferring the asymmetry from the standard model sector to the dark matter sector.

The total baryon number is conserved above the symmetry breaking scale.  Therefore the requirement of $B_T=0$ gives
\begin{align}\nonumber
B_T &= \frac{15}{ 4 \pi^2 g_{*} T} \Big[ 3 ( \mu_{u_L} +  \mu_{u_R} +  \mu_{d_L}  + \mu_{d_R}) \\
&\enspace + B_1 ( 2  \mu_{\Psi_L} +  \mu_{\eta_R} + \mu_{\chi_R}  ) +  B_2 (  2  \mu_{\Psi_R} +  \mu_{\eta_L} + \mu_{\chi_L} ) \nonumber \\
&\qquad - 6  \mu_{S_B}) \Big]=0 ,
\end{align}
\begin{figure}[b] 
   \includegraphics[width=3.2in]{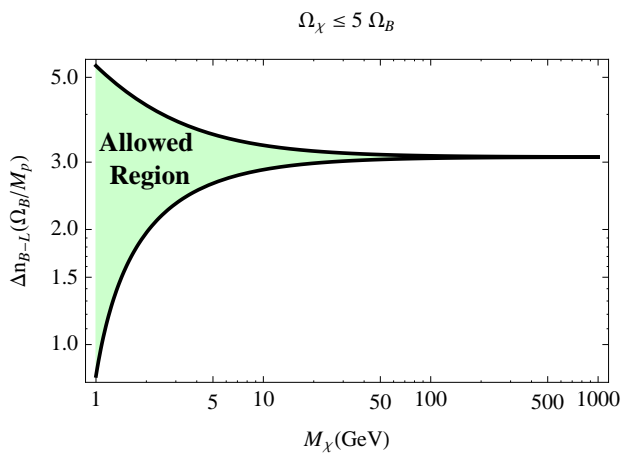} 
   \caption{Bounds on the initial asymmetry $\Delta n_{B-L}$ in units of $\Omega_B$ per proton mass, which satisfies the upper bound on the dark matter relic density Eq.~(\ref{eq:bound}), for the choice $B_2=-1$.}
   \label{fig:DM}
\end{figure}
Putting together the above relations, we find the following system of equations satisfied by the chemical potentials:
\begin{gather}\label{eq:system}
\begin{aligned}
\Delta(B\!-\!L)_{\text{SM}}&=\frac{15}{ 4 \pi^2 g_{*} T}(12 \mu_{u_L}-9\mu_{e_L}+3\mu_0),\\
\Delta\eta&=\frac{15}{ 4 \pi^2 g_{*} T} (4 \mu_{\Psi_L}+4 \mu_{\Psi_R}), \\
0&= 6 \mu_{u_L} + (2 B_1-3) \mu_{\Psi_L} + (2 B_2 + 3) \mu_{\Psi_R},  \\
0&= 3 \mu_{u_L} + 8 \mu_{0} - 3 \mu_{e_L} - \mu_{\Psi_L} - \mu_{\Psi_R}, \\ 
0 &= 9 \mu_{u_L} + 3 \mu_{e_L} + \mu_{\Psi_L} - \mu_{\Psi_R}.
\end{aligned}
\end{gather}
Now, the system of equations above can be solved to yield the final baryon asymmetry in the SM sector in terms of $\Delta(B-L)_{\text{SM}}$ and $\Delta\eta$ to yield
\begin{eqnarray}\nonumber
B^\text{SM}_f &\equiv& \frac{15}{ 4 \pi^2 g_{*} T}    \left( 12  \mu_{u_L} \right)\\
&=& C_1\,\Delta (B-L)_\text{SM} + C_2\,\Delta \eta,
\end{eqnarray}
where we find the values
\begin{equation}
C_1=\frac{32}{99} \qquad {}\text{and} \qquad C_2 = \frac{(15 - 14 B_2)}{198}.
\end{equation}
Here we have used the anomaly-free condition $B_1 = -3 + B_2$.  
Requiring that the dark matter asymmetry is bounded from above by the observed dark matter density, 
\begin{equation}
| n_\chi - n_{\bar{\chi}}| \leq n_{DM}\,,
\end{equation}
we find the following upper bound on the dark matter mass:
\begin{equation}\label{eq:bound}
M_\chi \leq \frac{\Omega_{DM} C_2 M_p}{ | \Omega_B - C_1 M_p \Delta n_{B-L}|}.
\end{equation}
Here $M_p$ is the proton mass and $\Delta n_{B-L}=s \Delta (B-L)_{SM}$. As one can appreciate, this bound is a function of the baryon number $B_2$, and the $B-L$ asymmetry generated through a mechanism such as leptogenesis. Therefore, in general one could say that we can have a consistent relation between the baryon and dark matter asymmetries.  Because the denominator can be small, the RHS could be large allowing for a large dark matter mass.  In Fig.~(\ref{fig:DM}), we display the allowed values for $\Delta n_{B-L}$ for the choice $B_2=-1$.

In order that the thermal component of dark matter does not oversaturate the observed density, one must understand how it can be minimized by considering annihilation mechanisms.  This issue has been studied in detail in Ref.~\cite{Duerr}.  In the case when the dark matter is light ($M_\chi<500$ GeV), one needs to postulate a new annihilation channel assuming, for example, a light extra gauge boson which mixes with the $U(1)_Y$ gauge boson \cite{Zurek}. When the dark matter is heavy ($M_\chi>500$ GeV), the thermal component of the dark matter relic density can be small due to the annihilation into two standard model quarks through the interaction with the leptophobic gauge boson $Z_B$.  As investigated in Ref.~\cite{Duerr}, one can have a large cross section in agreement with the constraints from direct detection.

We now briefly discuss scenarios 1 and 2 mentioned in the previous section.  In the case of scenario 1, where the leptophobic gauge boson mass is generated through the Stueckelberg mechanism, the extra fermions get mass only from the SM Higgs. We repeat the same analysis and find the connection between the baryon asymmetry and the dark matter density where the values of $C_1$ and $C_2$ are
\begin{equation}
C_1^\text{St}=\frac{16}{53} \qquad {}\text{and} \qquad C_2^\text{St} = \frac{(9 - 7 B_2)}{53}.
\end{equation}
The bound in Eq.~(\ref{eq:bound}) is relaxed due to the larger $C_2$ for a given value of $B_2$, allowing for heavier dark matter.  If we study the case in scenario 2, where $S_B$ carries charge $B\neq \pm3$, we find the same relationship between baryon asymmetry and dark matter density as in the Stueckelberg case.  Unfortunately, in these two cases the dark matter cannot be SM singlet-like because one does not have vector-like mass terms and it is difficult to satisfy bounds from dark matter direct detection experiments. \\
\section{Concluding Remarks}
We have proposed a new mechanism to simultaneously generate the baryon and dark matter asymmetries through the electroweak sphalerons.
This is realized in theories with low scale spontaneously broken gauged baryon symmetry.   
This mechanism provides an interesting theoretical framework to relate the baryon asymmetry and the dark matter asymmetric density in the universe.  We have investigated this issue in different scenarios corresponding to different schemes for the generation of mass for the leptophobic gauge boson. 

In these scenarios the $B-L$ asymmetry generated through a mechanism such as leptogenesis is transferred to the dark matter sector through sphaleron processes which conserve total baryon number. We have shown that it is possible to have a consistent relation between dark matter relic density and baryon asymmetry in the universe even if the baryon number is broken at the low scale.  We refer to this class of models as ``Asymmetric Baryonic Dark Matter''.

These results can be applied to more general theories where lepton number is also gauged, as proposed in Refs.~\cite{paper1,paper2,paper3}. In this case one can assume that the lepton number is broken at the high scale and a $B-L$ asymmetry is generated through leptogenesis.

{\textit{Acknowledgments:}}
{\small{P.F.P. thanks M.B. Wise for discussions.}}


\end{document}